\documentclass[twoside]{dis04}

\usepackage{amssymb}
\usepackage{amsmath}
\usepackage{epsfig}

\def\runauthor{Cyrille Marquet and Robi Peschanski}
\def\shorttitle{Saturation in two-hard-scale processes}

\newcommand{\be}{\begin{equation}}
\newcommand{\ee}{\end{equation}}
\newcommand{\bea}{\begin{eqnarray}}
\newcommand{\eea}{\end{eqnarray}}

\newcommand{\g}{\gamma}
\newcommand{\f}{\frac}

\newcommand{\intc}[1]{{\int\frac{d#1}{2i\pi}}}
\newcommand\lr[1]{{\left({#1}\right)}}

\begin{document}

\title{Saturation in two-hard-scale processes \\ at hadron colliders}

\author{Cyrille Marquet}

\address{Service de Physique Th\'eorique, CEA/Saclay\\
91191 Gif-sur-Yvette cedex, France\\
E-mail: marquet@spht.saclay.cea.fr}

\maketitle

\abstracts{A study of saturation effects in two-hard-scale hadronic processes 
such as Mueller-Navelet jets is presented. The cross-sections are expressed in 
the dipole framework while saturation is implemented via an extention of the 
Golec-Biernat and W\"usthoff model. The transition to saturation is found to be 
more abrupt than in $\gamma^*\!-\!\gamma^*$  cross-sections. Observables with a 
potentially clear saturation signal are proposed.}

\section{Introduction} 

Hard processes involving two perturbative scales lead to cross-sections whose 
linear high-energy behavior is described by the well-known 
Balitsky-Fadin-Kuraev-Lipatov (BFKL)  \cite{bfkl} equation. However, to 
respect whatever constraints unitarity may impose, it is well-known that the 
BFKL equation has to be modified beyond some energy limit, in order to describe 
cross-sections that saturate. Physically, the idea is that the gluon density in 
the BFKL ladder grows higher as one increases the energy and that eventually 
recombinations will occur, limiting the number of gluons in the ladder.

Three measurements for studying this behavior can be considered: the 
$\g^*\!-\!\g^*$ total cross-section in $e^+e^-$ scattering  \cite{gamma,sam}, 
Mueller-Navelet jets in hadron-hadron collisions  \cite{mnj}, and forward jets 
in 
deep inelastic scattering  \cite{dis,theo}. The perturbative scales in these 
processes are set by either the virtualities of the reaction-initiating photons 
or the transverse momenta of the measured jets. The aim of this work is to 
describe in a simple way how saturation effects could appear in those processes. 

Following the approach of Golec-Biernat and W\"usthoff whose saturation 
model  \cite{golec} (GBW) for the proton structure functions provides a simple 
and 
elegant formulation of the transition to saturation, we will implement 
saturation effects in the dipole framework  \cite{dipole,mueller}. The basis of 
this approach is to consider that the incident particules 
fluctuate into colorless quark-antiquark pairs (dipoles) which then interact. 
Saturation will then be modeled through the dipole-dipole scattering.

While such a study has already been done for the $\g^*\!-\!\g^*$ cross-section 
 \cite{motyka}, our work will focus on Mueller-Navelet jets; the extension to 
the 
forward-jet case will be straightforward. The key difference between the 
$\g^*\!-\!\g^*$ and the Mueller-Navelet jet measurements is that the hard probes 
are no more virtual photons but the final-state jets. The functions expressing 
the fluctuation of a virtual photon into dipoles are know from QED, but the 
descrition of a forward jet in terms of dipoles requires more care.
A first part is devoted to this problem and then saturation predictions within 
the GBW model are presented. Observables to be studied are proposed.

\section{Formulation} 

Mueller-Navelet jets are processes in which a proton strongly interacts with 
another proton or antiproton and where a jet with transverse momemtum larger 
than a perturbative scale is detected in each of the two forward directions. 
Such hard processes obey the collinear factorization which allows one to deal 
only with hard cross-sections. The two cuts on the jets transverse momenta will 
be denoted $Q_1$ and $Q_2$ and taken of the same magnitude in order to suppress 
the DGLAP evolution in the gluon ladder. The rapidity interval between the two 
jets $\Delta\eta$ is taken to be large in order to lie in the high-energy 
regime. 

Considering first the leading-logarithmic approximation when the evolution is 
linear, the dipole formulation of this hard total 
cross-section reads:
\begin{equation}
\sigma(Q_1,Q_2,\Delta\eta)=\int d^2r_1dz_1\ d^2r_2dz_2\ \phi_J(r_1,z_1,Q^2_1)\ 
\phi_J(r_2,z_2,Q^2_2)\ \sigma^{(0)}_{dd}(r_1,r_2,\Delta\eta)\ ,
\label{sigma}
\end{equation}
where $r_{i=1,2}$ are the transverse sizes of the dipoles interacting and 
$z_{i=1,2}$ are the fractions of longitudinal momentum of the quarks in each 
dipole. The BFKL dipole-dipole cross-section is
\begin{equation}
\sigma^{(0)}_{dd}(r_1,r_2,\Delta\eta)= \pi\alpha_s^2r_1^2
\intc{\g}\f{\lr{r_2/r_1}^{2\g}}{\g^2(1\!-\!\g)^2}\exp{\left\{\f{\alpha_s 
N_c}{\pi}\Delta\eta\ 
\lr{2\psi(1)\!-\!\psi(1\!-\!\g)\!-\!\psi(\g)}\right\}}\label{sdd}
\end{equation}
where $\psi(\g)$ is the logarithmic derivative of the 
Gamma function. The dipole distributions describing the forward-jet emissions 
have been denoted $\phi_J(r_i,z_i,Q^2_i).$

Let us recall how one can obtain this dipole distribution. The 
$k_T$-factorization property  \cite{kT} provides the general formalism for 
coupling external sources to the BFKL kernel through the convolution of impact 
factors. It can be proved that $k_T$-factorization is 
equivalent \cite{equivalent} to the dipole factorization expressed by formula 
(\ref{sigma}). The dipole distribution $\phi_J$ can thus be derived from the 
corresponding impact factors: the derivation \cite{robi,munier,us} is 
made using the example of a final-state gluon with transverse momentum larger 
than $Q$ being emitted off a perturbative onium ($q\bar q$ state) of size 
$r_0\!\ll\!1/\Lambda_{QCD}.$    
QCD factorization will allow to extend the result to the case of an incident 
hadron since the onium structure function factorizes out. Using 
$k_T$-factorization in the BFKL framework,
the impact factor $f(k^2,r_0)$ of the onium+jet system is related to the 
elementary gluon-dipole coupling 
$f^0(k^2,r)$ in the following way \cite{munier,us}:
\begin{equation}
f(k^2,r_0)=\left\{\f{2\alpha_s N_c}{\pi}\log\frac 1{x_J}
\log{Qr_0}\right\}
\int d^2r\  \frac Q{2\pi r}\ J_1(Qr)\ f^0(k^2,r) 
\label{impf}
\end{equation}
in the collinear approximation $Qr_0\!\gg\!1$ for the onium. $k$ is the 
transverse momentum of the gluon connected to the BFKL kernel and $x_J$ is the 
fraction of longitudinal momentum of the jet with respect to the onium. Formula 
(\ref{impf}) can be interpreted as the equivalence for forward jets between the 
momentum-space (partonic) and coordinate-space (dipole) representations. The 
factor in brackets $\left\{(2\alpha_s N_c/\pi)\log{Qr_0}\log 1/x_J\right\}$
corresponds to the probability of 
finding a dipole of size $1/Q$ inside the onium of size $r_0;$ thanks to QCD 
factorization properties, it is included in the gluon structure function of the 
incident particule (here the onium). $f^0(k^2,r)=(1\!-\!J_0(kr))/k^2$ is 
nothing 
else than the gluon density inside the dipole of size $r$ and, in the 
dipole formulation (\ref{sigma}), is included in the dipole-dipole 
cross-section $\sigma^{(0)}_{dd}$. Having factorized out both the contribution 
to the structure 
function and to the dipole-dipole cross-section, one is left with the function 
$\phi_J(r,Q^2)\equiv\int dz\ \phi_J(r,z,Q^2)$ which describes the resulting size 
distribution of the interacting dipole. Hence, one is led to identify 
\begin{equation}
\phi_J(r,Q^2)=\frac Q{2\pi r}\ J_1(Qr)\ .\label{phiJ}
\end{equation}

Let us now consider saturation effects. Initially, the GBW approach \cite{golec} 
is a model for the dipole-proton cross-section which includes the saturation 
damping of large-dipole configurations. For the description of $\g^*\!-\!\g^*$ 
cross-sections at LEP  \cite{motyka}, it has been extended to 
dipole-dipole cross-sections: 
\begin{equation}
\sigma_{dd}(r_1,r_2,\Delta\eta)=\sigma_0\left\{
1\!-\!\exp\lr{-\f{r_{\rm eff}^2(r_1,r_2)}{4R_0^2(\Delta \eta)}}\right\}\ .
\label{sigmadd}
\end{equation}
The dipole-dipole {\it effective} radius $r^2_{\rm  eff}(r_1,r_2)$ is defined 
through the two-gluon exchange:
\be 2\pi\alpha_s^2r^2_{\rm eff}(r_1,r_2)\equiv\sigma^{(0)}_{dd}(r_1,r_2,0)
=2\pi\alpha_s^2\min(r_1^2,r_2^2)\left\{1\!+\!\log\frac{\max(r_1,r_2)}
{\min(r_1,r_2)}\right\}\label{reff}
\ee
while for the saturation radius 
$R_0(\Delta\eta)\!=\!e^{-\f{\lambda}2\left({\Delta 
\eta}-{\Delta \eta}_0\right)}/Q_0$ we shall use the same set of parameters as 
those in \cite{golec,motyka}, that is $\lambda\!=\!0.288,$ $\Delta 
\eta_0\!=\!8.1$ for $Q_0\!\equiv\!1\ GeV.$
Two other scenarios for $r_{\rm eff}(r_1,r_2)$ have also been considered: 
$r^2_{\rm  eff}\!=\!r_1^2r_2^2/(r_1^2+r_2^2)$ and $r^2_{\rm 
eff}\!=\!\min(r_1^2,r_2^2).$

We shall use $\sigma_{dd}$ in the hard cross-section (\ref{sigma}) instead of 
$\sigma^{(0)}_{dd}$ to implement saturation in a simple way. However, in order 
to do 
so, one makes the non-trivial assumption that the dipole factorization still 
holds when the dipole-dipole cross-section is modified by saturation.

\section{Phenomenology} 

Inserting (\ref{phiJ}) and (\ref{sigmadd}) in formula (\ref{sigma}) leads to the 
simple final result for the Mueller-Navelet hard cross-sections modified by 
saturation within the GBW model:
\be
\sigma(Q_1,Q_2,\Delta\eta)/\sigma_0=1-2R^2_0Q_1Q_2\int_0^\infty du\
\f{e^{-(Q^2_1+Q^2_2u^2)R^2_0/r^2_{\rm eff}(1,u)}}{r^2_{\rm 
eff}(1,u)}I_1\left(\frac{2Q_1Q_2uR^2_0}
{r^2_{\rm eff}(1,u)}\right)\ .
\label{final}
\ee
Some comments are in order. The dipole distribution 
$\phi_J(r,Q^2)$ is not everywhere positive and we interpret this feature as a 
breakdown of the collinear approximation. It also means that one has to check 
that replacing $\sigma^{(0)}_{dd}$ by $\sigma_{dd}$ in (\ref{sigma}) does not 
alter 
the positivity of the hard cross-sections, and this is indeed the case. Another 
check that our approximations require is that the cross-sections 
$\sigma_{dd}\!\sim\!\sigma_0{r_{\rm eff}^2/4R_0^2(\Delta \eta)},$ corresponding 
to 
the limit of small dipole sizes in (\ref{sigmadd}), lead to hard cross-sections 
behaving like $1/\left\{R_0^2(\Delta\eta)\max{(Q_1^2,Q_2^2)}\right\},$ as 
expected from transparency. The model $r^2_{\rm eff}\!=\!\min(r_1^2,r_2^2)$ 
does 
not and therefore we cannot consider it in our approximations.

Let us investigate the phenomenological outcome, for hadron colliders, of our 
extension of the GBW models to Mueller-Navelet jets. The theoretical hard 
cross-sections are obtained from formula (\ref{final}) in terms of the physical 
variables $Q_1$, $Q_2$ and $\Delta\eta.$ When plotting them, one observes the 
expected trend of the GBW model, that is a convergence of the cross-sections  
towards the full saturation limit $\sigma\rightarrow\sigma_0.$ In order to 
appreciate more quantitatively the influence of saturation, it is 
most convenient to consider the quantities ${\cal R}_{i/j}$ defined as 
\begin{equation}
{\cal R}_{i/j}\equiv  
\frac {\sigma(Q_1,Q_2,{\Delta \eta}_i)}{\sigma(Q_1,Q_2,{\Delta \eta}_j)}\ , 
\label{Rij}
\end{equation}
{\it i.e.} the cross-section ratios for two 
different values of the rapidity interval. These ratios 
display in a clear way the saturation effects. They also correspond to possible 
experimental observables since they can be obtained from measurements at  
fixed values of the jets longitudinal momenta $x_{J_1}$ and $x_{J_2}$ and thus 
are independent of the structure functions $f(x_{J_i},Q^2_i)$ of the incident 
hadrons. Indeed, the experimental measurement is
\be
\f{d\sigma_{tot}^{pp\rightarrow jj+X}}{dx_{J_1}dx_{J_2}}
=f(x_{J_1},Q^2_1)f(x_{J_2},Q^2_2)\ \sigma(Q_1,Q_2,\Delta\eta)
\ee
and the ratio of these cross-sections gives access to ${\cal R}$. Such 
observables have actually been used for a study of Mueller-Navelet jets for 
testing BFKL predictions at the Tevatron  \cite{goussiou,jets}. 

In Fig.1 we plot 
the values of ${\cal R}_{4.6/2.4}$ (resp. ${\cal R}_{8/4}$) as a function of 
$Q_1\!=\!Q_2\!\equiv\!Q.$ ${\cal R}_{4.6/2.4}$ is the observable that has been 
considered for the Tevatron \cite{goussiou,jets} while ${\cal R}_{8/4}$ 
corresponds to realistic rapidity intervals for the LHC. As 
expected from the larger rapidity range, the decrease of ${\cal R}$ between the 
transparency regime and the saturated one is larger for the LHC than for the 
Tevatron. The striking feature of Fig.1 is that the effect of saturation appears 
as a sharp transition for some critical range $Q\!\sim\!1/R_0.$ No saturation 
effects would correspond to ${\cal R}$ constant equal to the high$-Q^2$ limit of 
the plots while the full-saturated limit is ${\cal R}\!=\!1.$ Comparing these 
ratios for Mueller-Navelet jets to 
those for the $\g^*\!-\!\g^*$ measurement for the same values of the 
rapidity ranges, one interestingly sees that the $\g^*\!-\!\g^*$ transition 
curve is much smoother, a phenomenon explained by the different structure 
of the dipole distributions. Indeed the formula to compute the $\g^*\!-\!\g^*$ 
case is also formula (\ref{sigma}) but with of course the well-known photon 
dipole 
distributions $\phi_\g$ instead of $\phi_J.$ As discussed in  \cite{robi}, the 
dipole distribution $\phi_J(r,Q^2)$ has a tail extending towards large dipole 
sizes, which are more damped by the saturation corrections. Hence $\phi_J$ is 
more abruptly cut by saturation than the photon dipole distribution $\phi_\g$. 
Note that saturation effects in forward-jets  \cite{fjets} can be studied in a 
straightforward manner using our formalism: it requires to combine (\ref{sigma}) 
with both dipole distributions $\phi_\g$ and $\phi_J.$

The signal displayed in Fig.1 shows a clear transition to saturation, however 
the values of $Q$ at which it occurs are rather low, probably to low for 
experimental $E_T\!-\!cuts$ on jets. An interesting way out of this problem 
could be 
that the saturation scale is higher than the one we used in the present work, 
namely the one extracted from $F_2.$ Indeed, it has been proposed \cite{levin} 
that the saturation scale could be higher for two-hard-scale processes like 
Mueller-Navelet jets than for one-hard-scale measurements like the proton 
structure functions. That would shift the transition shown in Fig.1 
towards higher $Q.$ 
\begin{figure}
\begin{center}
\epsfig{file=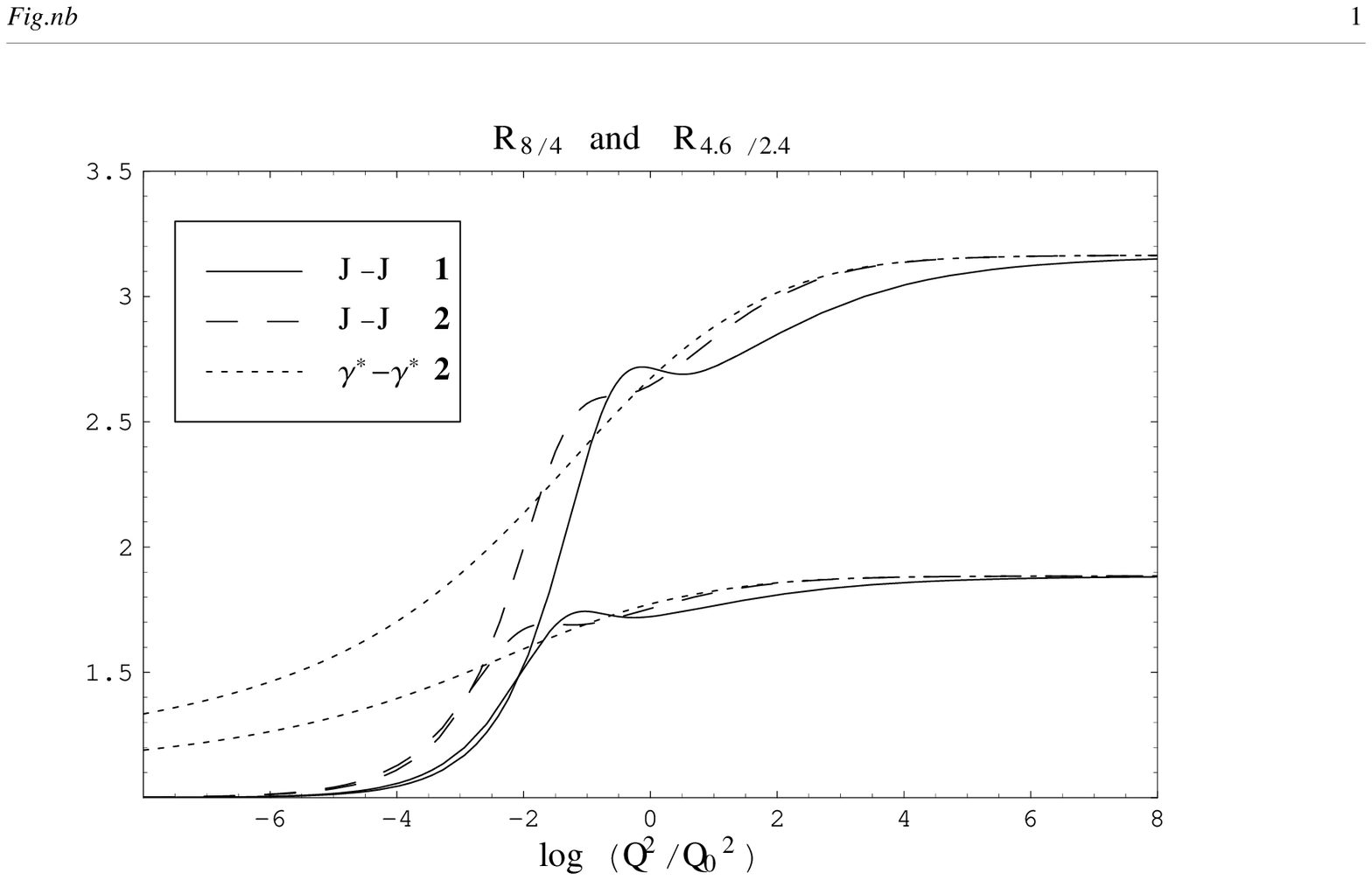,width=14cm,clip=true}
\caption{{\it Cross-section ratios ${\cal R}_{i/j}.$} The resulting ratios 
for the two-gluon exchange model ({\bf 1}) and for $r^2_{\rm  
eff}\!=\!r_1^2r_2^2/(r_1^2+r_2^2)$ ({\bf 2}) are plotted for rapidity intervals 
$i\!=\!8,j\!=\!4$ (highest set of curves) 
and $i\!=\!4.6,j\!=\!2.4$ (lowest set of curves). The comparison is made with 
$\gamma^*\!-\!\gamma^*$ ratios 
for model {\bf 2} and equivalent kinematics.}
\end{center}
\label{F}
\end{figure}
Another alternative to solve this ``low-Q'' problem would be to consider the 
detection of heavy vector or heavy-flavored mesons as alternatives to forward 
jets. Indeed, using $J/\Psi's,$ $\Upsilon's,$ ${D^*}'s,$ or $B-$mesons may 
provide 
hard probes of lower transverse momenta than jets, allowing to look deeper 
in the saturation regime.

These possibilities of realizing hard hadronic probes 
of saturation certainly deserve more studies in the near future. On the 
theoretical side, going beyong our approximations seems necessary while on the 
phenomenological side, simulations at 
Tevatron and LHC energies will be needed to give a quantitative estimate of the 
potential of hadronic colliders to reveal those new features of saturation.

\section*{Acknowledgements} 
The author wishes to thank Robi Peschanski for the collaboration. He also thanks 
St\'ephane Munier and Chritophe Royon for useful comments and suggestions.

\end{document}